\documentclass[twocolumn,showpacs,preprintnumbers,amsmath,amssymb,prl]{revtex4} 

\usepackage{graphicx}
\usepackage{dcolumn}
\usepackage{bm}

\begin{document}

\title{Coarse grained Density Functional theory of order-disorder phase transitions in metallic alloys}

\date{\today}

\author{Ezio Bruno}
\email{ebruno@unime.it}
\author{Francesco Mammano}
\author{Beniamino Ginatempo}

\affiliation{ Dipartimento di Fisica,
Universit{\`{a}} di Messina, Salita Sperone 31, 98166 Messina, Italy.}


\begin{abstract}
The technological performances of metallic compounds are largely 
influenced by atomic ordering. Although there is a general consensus 
that successful theories of metallic systems should account for the 
quantum nature of the electronic glue, existing non-perturbative high-temperature 
treatments are based on effective classical atomic Hamiltonians. 
We propose a solution for the above paradox and offer a fully
quantum mechanical, though approximate, theory that 
on equal footing deals with both electrons and ions. By taking advantage of
a coarse grained formulation of the density functional theory [Bruno et al.,
Phys. Rev. B {\bf 77}, 155108 (2008)] we develop a MonteCarlo technique, 
based on an \emph{ab initio}
Hamiltonian, that allows for the efficient evaluation of finite temperature
statistical averages. Calculations of 
the relevant thermodynamic quantities and of the electronic 
structures for CuZn and Ni$_3$V support that our theory provides 
an appropriate description of order-disorder phase transitions. 
\end{abstract}

\maketitle

Nowadays ground state properties of metallic alloys are routinely calculated by
using Density Functional (DF)~\cite{H&K,K&S} theory which appropriately incorporates electronic correlations. 
However, most finite temperature status of art calculations~\cite{Turchi} are based on classical Ising models which
allow the accurate evaluation of phase equilibria, but are not able to predict the electronic properties. 
A theory able to cope with both tasks must incorporate quantum mechanics and should be able
to explore the very large space of alloy configurations~\cite{Franceschetti} in order to give reliable statistical averages. 
Unfortunately, quantum simulations as the Car-Parrinello Molecular Dynamics~\cite{CarParrinello} (CPMD) are very hard 
for the problem at hand and the application of such methods the 
phase equilibria of metallic alloys seems beyond the capabilities of existing hardware and software.   
In this Letter we shall outline a new approach to the problem, which, similarly to CPMD is based on the 
DF theory and the Born-Oppenheimer approximation, but where the relevant quantities shall be obtained
as statistical averages (by sampling the space of alloy configurations) rather than as Boltzmann time averages.

Although the generalization to n-ary systems is straightforward, in this Letter we 
focus on binary metallic alloys A$_c$B$_{1-c}$. They shall be studied in the $(T,c)$ statistical 
ensemble defined by the temperature, $T$, and the mean atomic concentration, $c$. In 
order to have a tractable problem, we shall limit ourselves to the solid state and 
the normal metal regime or, equivalently, to the temperatures between the 
superconducting and the melting transitions, $T_{SC}<T<T_M$. Furthermore, martensitic and 
magnetic phenomena shall not be considered.

Our first crucial step is to obtain a \emph{coarse grained} version of the Hohenberg-Kohn 
DF. If the ions constituting 
the system are considered frozen on the  sites of a simple lattice at their 
equilibrium positions, $\mathbf{R}_i$, then the electrostatic contribution to the DF can be 
written as a sum of  \emph{local} terms plus some \emph{bilinear} terms involving the \emph{local 
moments} of the electronic density~\cite{BMFM,nota1}:
\begin{equation}\label{eq1}
U^{el}([\rho(\mathbf{r})]) = \sum_i  u_i([\rho_i(\mathbf{r})])  +  \frac{e^2}{2} \sum_{i,j}  q_i M_{ij} q_j	
\end{equation}
In Eq.~(\ref{eq1}), each lattice site is associated with a Voronoi polyhedron (VP), i.e. to 
the set of points closer to that particular rather than to any other lattice site, 
the $u_i([\rho_i(\mathbf{r})])$ are known functionals of the local electronic density, 
$\rho_i(\mathbf{r})$, $-e q_i$ are the charge multipole moments in each VP and the Madelung matrix, 
$\mathbf{M}$, is determined by the crystal geometry only. As commented in Ref.~\onlinecite{BMFM}, 
the coupling between different VP's in Eq.~(\ref{eq1}) is \emph{marginal}~\cite{BMFM}, in the sense 
that it has a simple, analytically 
tractable form. Most existing approximations for the exchange-correlation energy~\cite{Dreizler&Gross} 
also consist of sums of local electronic density functionals. However, the  
local kinetic contributions to the DF are non trivially 
entangled by the boundary conditions (bc) that the wave-function, or the Green 
function~\cite{Gonis}, must match at the VP surfaces. A marginally coupled  
DF then requires an approximation for the kinetic part. The literature reports several ways in which 
such decoupling has been obtained: Ref.~\onlinecite{Harris} uses vacuum bc's, while random bc's
are employed in Ref.~\onlinecite{Krajewski&Parrinello}. In our approach 
the 'true' bc's are replaced by appropriate mean field bc's~\cite{BMFM}. The same procedure 
defines the class of the Generalized Coherent Potential Approximations (GCPA), 
whose prototype, the Isomorphous CPA (ICPA), has been introduced by Soven~\cite{Soven}.  
The ICPA provides an excellent picture~\cite{Abrikosov_cpa} of the spectral properties of metallic 
alloys but may lead to incorrect predictions about the electrostatics~\cite{Magri}. Modern GCPA 
schemes like the Polymorphous CPA~\cite{Ujfalussy} (PCPA) heal the shortcomings of the old ICPA model 
and provide fairly good total energies both for ordered and disordered alloy 
configurations~\cite{BMFM}. We shall then use the GCPA DF~\cite{BMFM}:
\begin{eqnarray}\label{eq2}
&&\Omega^{GCPA}([\rho_i(\mathbf{r})],\mu;\{\xi\})  = \\ \nonumber &&\sum_i  \omega^\alpha([\rho_i(\mathbf{r})]) 
 +  \frac{e^2}{2} \sum_{ij} q_i \mathbf{M}_{ij} q_j - \mu \sum_i q_i	
\end{eqnarray}
In Eq.~(\ref{eq2}), the dependence on the alloy configuration is identified by the set of 
occupation numbers $\{\xi\}$, with $\xi_i=1$ (or $0$) for sites occupied by an A (or B)  atom. 
In the case of a simple lattice, the local functionals $\omega^\alpha([\rho_i(\mathbf{r})])$ parametrically 
depend on the site chemical occupation only, $\alpha=$~A or B. Minimizing Eq.~(\ref{eq2}) with 
respect to the $\rho_i(\mathbf{r})$ provides a set of  Euler-Lagrange equations coupled only 
through the Madelung potentials, $V_i =\sum_j M_{ij} q_j$, and the chemical potential $\mu$: 
\begin{equation}\label{eq3}
\frac{\delta \omega^\alpha}{\delta \rho_i(\mathbf{r})}  +  e^2 V_i = \mu
\end{equation}

Within the alloy sample specified by $\{\xi \}$, due to Eq.~(\ref{eq3}) and to the variational 
properties of the GCPA DF~\cite{BMFM,DFTKKRCPA1,DFTKKRCPA2}, the local 
charge densities for each chemical component, $\rho^\alpha(\mathbf{r})$, are \emph{unique functions}~\cite{BMFM} of the 
Madelung potentials. By virtue of the Hohenberg-Kohn theorem, the same holds for any 
other local physical observable,  
\begin{equation}\label{eq4}
\langle O\rangle_i=O^\alpha(V_i)
\end{equation}
Evidently, also the charge moments, $q_i$, are unique functions of the $V_i$. If, as it can 
be argued on a physical ground, the $qV$ laws, $q_i=q^\alpha(V_i)$, are strictly monotonic~\cite{BMFM}, 
then they can be inverted, thus allowing to recast the first term in Eq.~(\ref{eq2}) as a \emph{function} of the charge 
moments rather than a \emph{functional} of the full charge density, 
$\omega^\alpha([\rho_i(\mathbf{r})]) \rightarrow \tilde{\omega}^\alpha(q_i)$. This gives the desired 
coarse graining.

Actually, in the metallic state, the $qV$ laws are not only monotonic but numerically 
almost indistinguishable from straight lines~\cite{BMFM,FWS1,CEF}. Thus, an excellent approximation 
for Eq.~(\ref{eq2}) can be obtained by a series expansion about the zero field (or ICPA) 
values of the charge moments, $q_i^0$. This gives the Charge Excess Functional (CEF) 
already obtained on a phenomenological ground in Ref.~\cite{CEF},
\begin{eqnarray}\label{eq5}
&&\Omega^{CEF}(\{q\},\mu;\{\xi\})  = \\ \nonumber && \sum_i \frac{a_i}{2} (q_i-q_i^0)^2 
+ \frac{e^2}{2} \sum_{ij} q_i M_{ij} q{j} - \mu \sum_i q_i
\end{eqnarray}
where $a_i=(d^2 \tilde{\omega}^\alpha(q)/d q^2)_{q=q_0}$ takes the values $a_A$ or $a_B$, depending on the site occupation. 
For a given configuration, $\{\xi\}$, the minimum value of the functional in Eq.~(\ref{eq2}) or~(\ref{eq5}) 
corresponds to the total system energy and, therefore, provides an \emph{ab initio} effective 
atomic Hamiltonian~\cite{CEF}. Unlike most Ising models, such CEF Hamiltonian includes effective 
interactions at all distances and $n$-body interactions up to any value for $n$~\cite{brunomatsci}. For a 
binary alloy, it is determined by only three parameters that can be easily obtained by 
GCPA-DF calculations. 

\begin{figure}
\begin{center}
   \includegraphics[width=7cm]{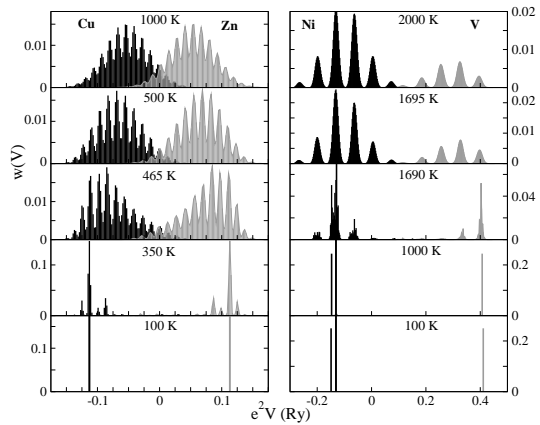}
\caption{Distributions of the Madelung potentials (DMP). 
Calculated DMP's for CuZn (left) and Ni$_3$V (right) alloys at the indicated temperatures. The structures visible in 
the DMP's are associated with the local atomic environments. Above the transition temperature ($T_O \approx 465 K$ for CuZn 
and $T_O= 1701\pm 11 K$ for Ni$_3$V) the DMP's present an overall Gaussian shape, while at lower temperatures 
they resemble the $T \rightarrow 0$ limit, consisting of $\delta$-like peaks, one for each non-equivalent lattice site.}
\label{fig1}
\end{center}
\end{figure}

\begin{figure}
\begin{center}
   \includegraphics[width=7.5cm]{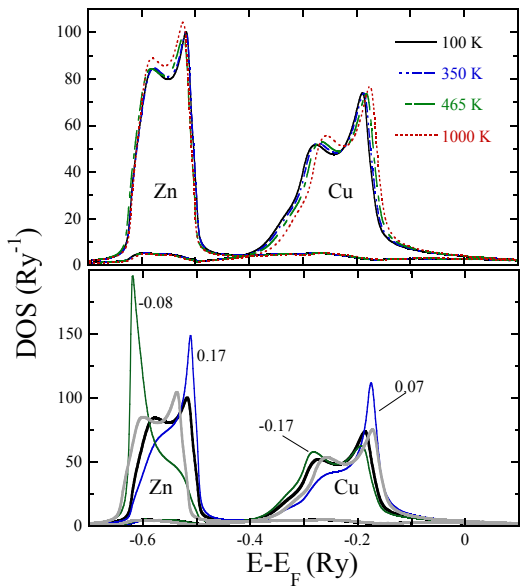}
\caption{Site resolved electronic DOS's for CuZn alloys. Upper frame: mean DOS contributions from Cu and 
Zn sites at the indicated temperatures. 
Lower frame: the comparison between the mean DOS's from Cu and Zn sites at $T=350 K$ 
(thick black lines) with those from the most extreme 
chemical environments in the GCPA ensemble (ligh lines, the corresponding values for 
$e^2V_i$ are indicated by the labels) 
shows the sensitivity of the GCPA theory to the site environments. The lower frame 
reports also the ICPA site resolved 
DOS evaluated at the same lattice constant (thick gray lines).
}
\label{fig2}
\end{center}
\end{figure}

\begin{figure}
\begin{center}
   \begin{tabular}{cc}
   \includegraphics[width=4cm]{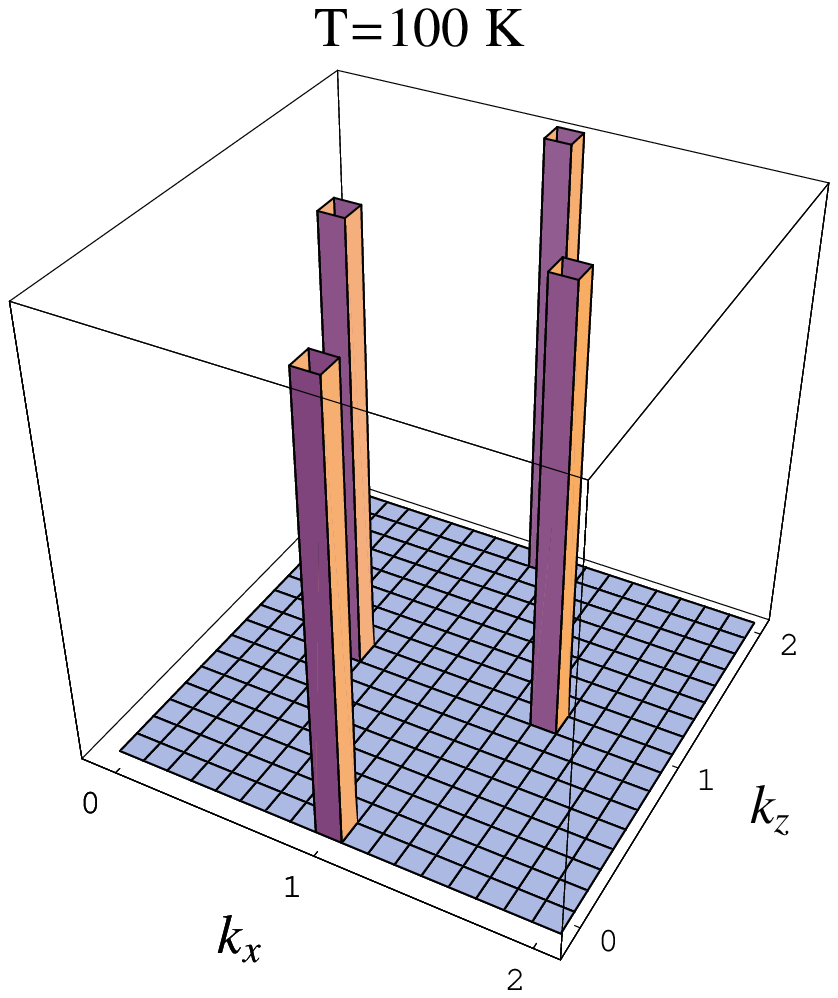} 
   \includegraphics[width=4cm]{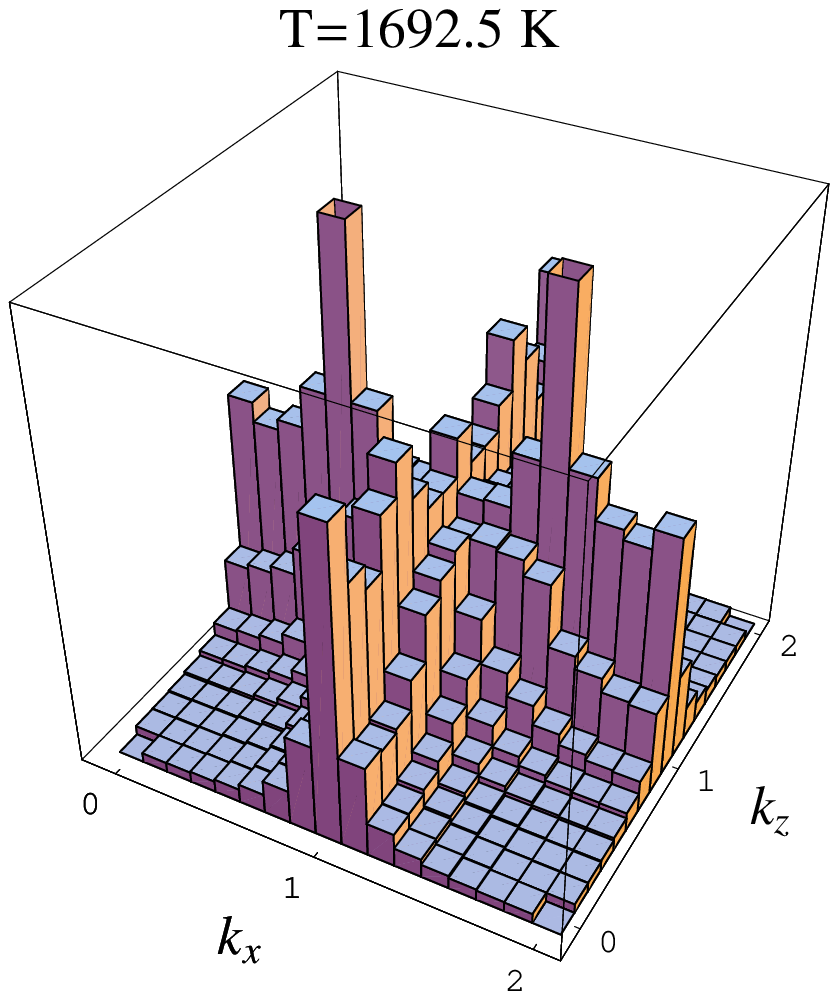}\\[0.5cm]
   \includegraphics[width=4cm]{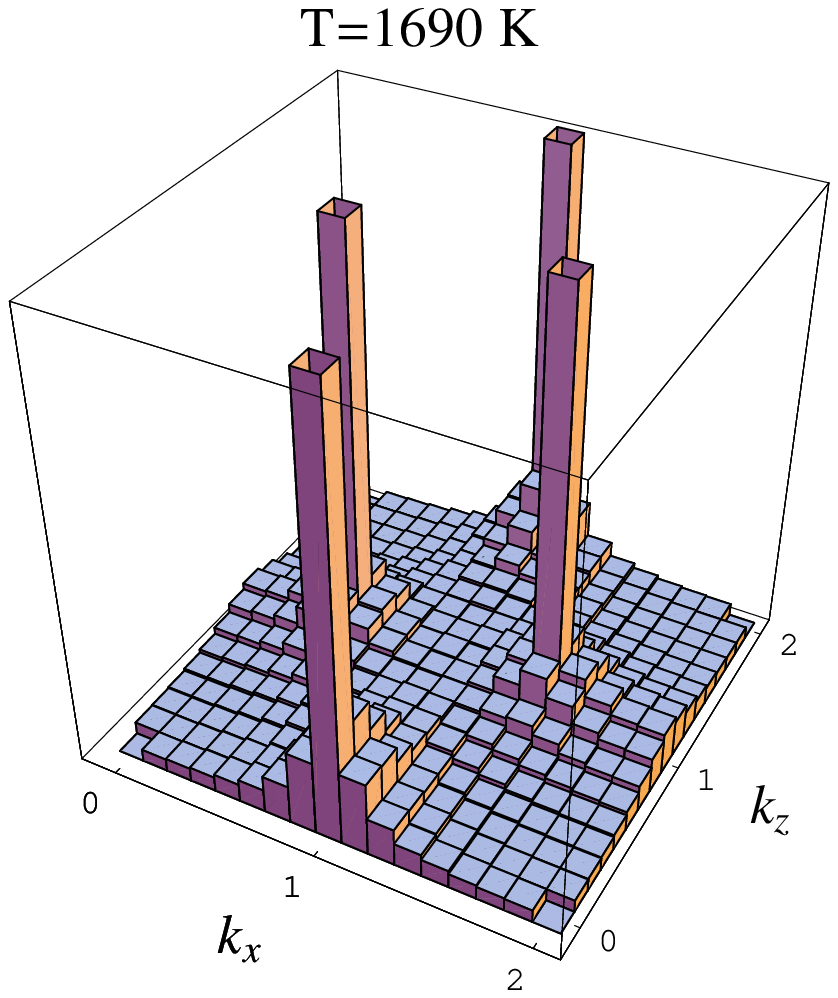}
   \includegraphics[width=4cm]{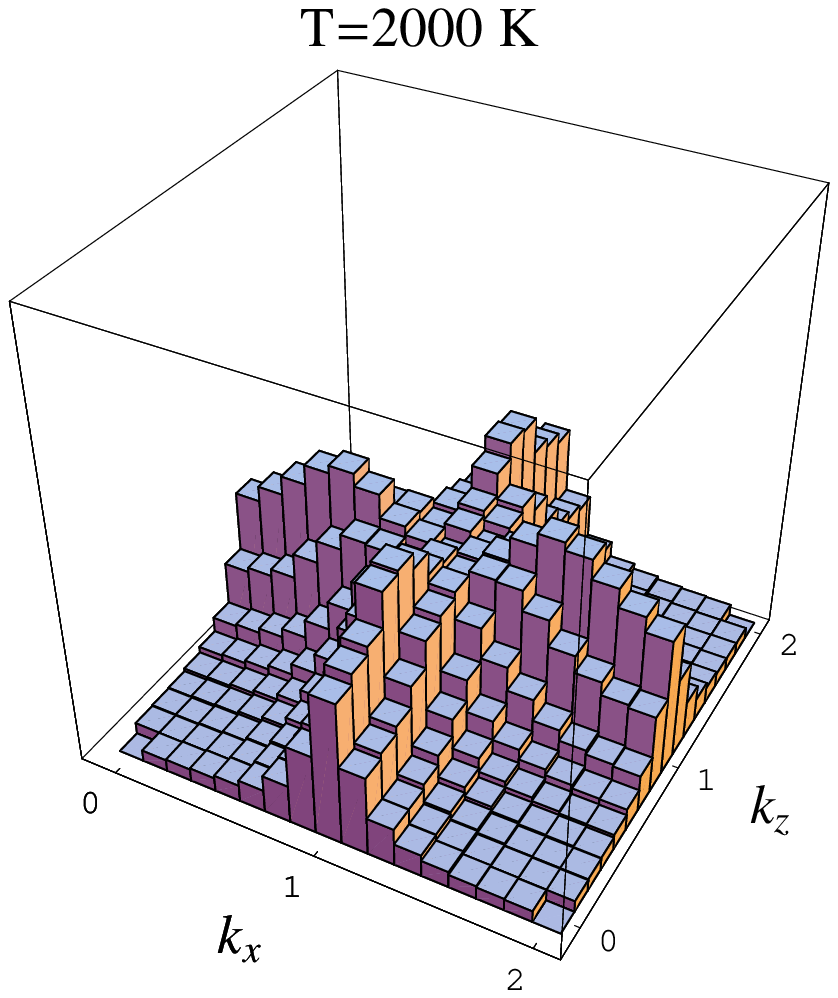}
\end{tabular}
\caption{Atomic Short Range Order (SRO) for Ni$_3$V alloys in the $k_y=0$ plane, at
the indicated temperatures. Above $T_O$, the $k_x$ and $k_z$ directions are equivalent, due to 
the cubic symmetry of the disordered fcc phase. Below $T_O$, the cubic symmetry is broken, giving rise to the tetragonal 
symmetry of the DO$_{22}$ structure, with the characteristic peaks at $(1,0,0)$ and $(1/2,0,1)$. It is remarkable the 
abrupt change of the structure~\cite{nota2}, that on cooling occurs between 1692.5~K 
and 1690~K.
}
\label{fig3}
\end{center}
\end{figure}

In this Letter we implement a coarse grained all electrons theory based on the GCPA 
DF and a finite temperature Monte Carlo (MC) sampling~\cite{Metropolis,Landau&Binder} of the alloy configurations 
space. This theory, in the following 
referred to as GCPA-CEF-MC, consists of three major stages~\cite{nota2}: 
(i) the GCPA DF is determined by $T=0$ electronic structure calculations~\cite{nota3}; 
(ii) for a given thermodynamic point $(T,c)$  the relevant ensemble of 
configurations is sampled by a MC based on the CEF Hamiltonian; for each 
configuration, the local charges and Madelung potentials are obtained by minimizing 
Eq.~(\ref{eq5}), whose solutions are numerically indistinguishable~\cite{BMFM} from those of Eq.~(\ref{eq2}); 
(iii) the obtained distribution of the Madelung potentials (DMP) (see Fig.~\ref{fig1}) then 
allows for the evaluation of the appropriate ensemble averages for the electronic 
properties through Eq.~(\ref{eq4}). As an example, we report in Fig.~\ref{fig2} the electronic density 
of states (DOS) for CuZn alloys at several temperatures. The stage (ii) of our 
GCPA-CEF-MC provides the ensemble averages corresponding to the relevant thermodynamic 
quantities and to the properties related with the atomic degrees of freedom only. 
Such is, for instance, the atomic Short Range Order (SRO) (see Fig.~\ref{fig3}, where results 
for Ni$_3$V alloys are presented). 

\begin{figure}
\begin{center}
   \includegraphics[width=7cm]{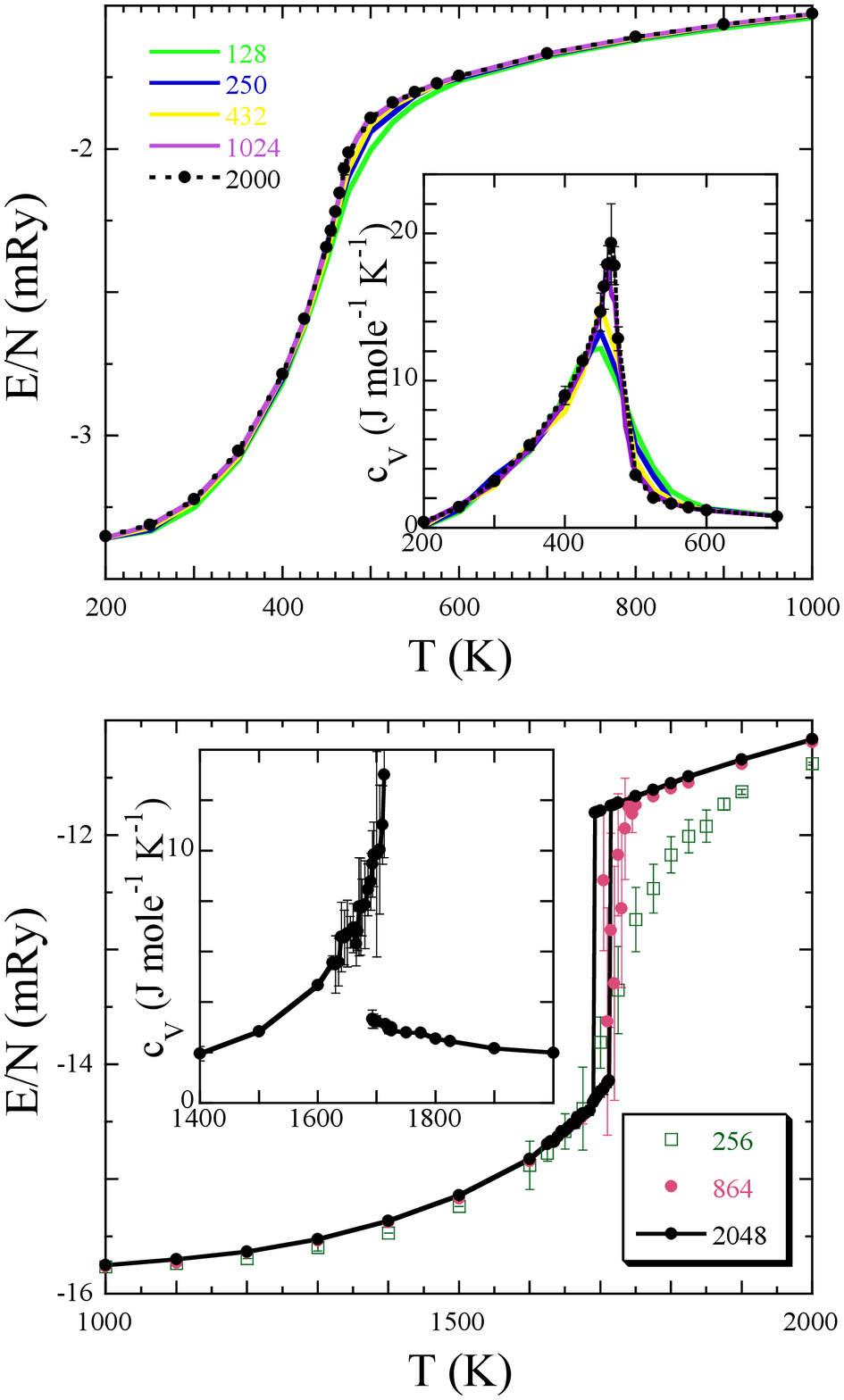} 
\caption{Electronic contributions to the total energies, $E$, and the specific heats, 
$c_V$, for CuZn (upper frame) and Ni$_3$V (lower frame) alloys 
as a function of the temperature.
The specific heats (insets) identify the ordering temperatures, $T_O$, with the respective uncertainties indicated by the 
shaded regions, and exhibit the typical $\lambda$-shape. Different symbols refer to different sizes ($N$) of the simulation 
boxes. CuZn energies do not show hysteresis, as expected for the second order transition bcc-B$_2$. For the largest 
simulation box, Ni$_3$V energies show a hysteresis loop, revealing the first-order character of the symmetry breaking  
fcc-DO$_{22}$ transition~\cite{nota2}.
}
\label{fig4}
\end{center}
\end{figure}

In order to test our GCPA-CEF-MC method, we have selected two well studied systems: 
CuZn~\cite{alphabrass,optcuzn,Tulip}, already discussed by Landau~\cite{LLstphysI} as the prototype for his theory of 
second-order phase transitions, and Ni$_3$V~\cite{Johnson2000,Zarkevich2004}, for which there is competition between 
different ordered structures. For both alloys thermal expansion has been included~\cite{DebyeGruneisen}.
This inclusion, however, implies only small quantitative changes to our results.   
Below $T_M$, both systems present disordered solid solution phases, based on the 
relatively open bcc lattice for CuZn and on the close-packed fcc for Ni$_3$V. On 
decreasing $T$, both systems undergo a transition to the ordered phases, B$_2$ and DO$_{22}$ 
respectively. The phase transitions have been monitored by plotting the total energies 
and the specific heats as a function of $T$ (Fig.~\ref{fig4}). The calculations for CuZn alloys 
show the neat evidence of an order-disorder transition occurring at $T_O \approx 465 K$, with a 
smooth dependence on N, the number of atoms contained in the simulation box. In the 
vicinity of the transition, quite large N values are necessary to obtain well converged 
thermodynamic properties for the Ni$_3$V system. For the largest simulation box, ordering 
occurs on cooling between 1692.5 and 1690 K, while, on heating, the low-$T$ ordered state 
disorders between 1710 and 1712.5 K. Our best estimate for $T_O$ in Ni$_3$V is then $1701 \pm 11$ K. 
The transition reveals a first-order character, as confirmed by the abrupt change in 
the SRO's displayed in Fig.~\ref{fig3}. The sharp first-order features of the transformation 
are gradually suppressed by finite size effects in the smaller simulation boxes investigated.

Our results demonstrate that the GCPA-CEF-MC theory is able to provide a sensible 
description of ordering phenomena in metallic alloys. Remarkably, we find complete 
agreement with the experiment about the low-$T$ ordered phases and the theory correctly 
discriminates between first- and second-order transitions. The agreement with the 
experiment about the ordering temperatures, however, is not completely 
satisfactory. The experimental ordering temperatures~\cite{ASM} for CuZn and Ni$_3$V, 
740 K and 1313 K, respectively, are in one case lower and in the other higher than 
the calculated values. This 
is probably due to the single-site nature of the GCPA and to the approximations~\cite{nota3} made 
for the site potentials. 

The methodologies presented in this Letter offer a route to alloy thermodynamics that 
fully includes electrons and appreciably enlarges the scales at which quantum 
mechanics can be applied. The calculations here discussed consider up to 2000 
atoms and up to $10^{11}$ local chemical environments for each point in the $(T,c)$ space. 
It is clear that such impressive figures can be obtained only because we have a fixed 
crystal lattice. The application of our method is then complementary to the 
CPMD, which is very hard to apply to diffusive transitions in the solid state.  
We are confident that, in the next future, coarse grained DF theories 
shall be extended to magnetic metals and to the mesoscopic domain, in order to study 
defects and plasticity.

\begin{acknowledgments}
The calculations presented have been executed 
using the computational facilities located at the CINECA (Bologna, Italy) and 
at the Consorzio COMETA, project 
PI2S2 (http://www.pi2s2.it).
\end{acknowledgments}


\begin{thebibliography}{32}
\expandafter\ifx\csname natexlab\endcsname\relax\def\natexlab#1{#1}\fi
\expandafter\ifx\csname bibnamefont\endcsname\relax
  \def\bibnamefont#1{#1}\fi
\expandafter\ifx\csname bibfnamefont\endcsname\relax
  \def\bibfnamefont#1{#1}\fi
\expandafter\ifx\csname citenamefont\endcsname\relax
  \def\citenamefont#1{#1}\fi
\expandafter\ifx\csname url\endcsname\relax
  \def\url#1{\texttt{#1}}\fi
\expandafter\ifx\csname urlprefix\endcsname\relax\def\urlprefix{URL }\fi
\providecommand{\bibinfo}[2]{#2}
\providecommand{\eprint}[2][]{\url{#2}}

\bibitem[{\citenamefont{Hohenberg and Kohn}(1964)}]{H&K}
\bibinfo{author}{\bibfnamefont{P.}~\bibnamefont{Hohenberg}} \bibnamefont{and}
  \bibinfo{author}{\bibfnamefont{W.}~\bibnamefont{Kohn}},
  \bibinfo{journal}{Phys.\ Rev.} \textbf{\bibinfo{volume}{136}},
  \bibinfo{pages}{B864} (\bibinfo{year}{1964}).

\bibitem[{\citenamefont{Kohn and Sham}(1965)}]{K&S}
\bibinfo{author}{\bibfnamefont{W.}~\bibnamefont{Kohn}} \bibnamefont{and}
  \bibinfo{author}{\bibfnamefont{L.~J.} \bibnamefont{Sham}},
  \bibinfo{journal}{Phys.\ Rev.} \textbf{\bibinfo{volume}{140}},
  \bibinfo{pages}{A1133} (\bibinfo{year}{1965}).

\bibitem[{\citenamefont{Turchi et~al.}(2007)\citenamefont{Turchi, Abrikosov,
  Burton, Fries, Grimvalle, Kaufman, Korzhavyi, Manga, Ohno, Pisch
  et~al.}}]{Turchi}
\bibinfo{author}{\bibfnamefont{P.~E.~A.} \bibnamefont{Turchi}}~
  \bibnamefont{et~al.}, \bibinfo{journal}{Calphad}
  \textbf{\bibinfo{volume}{31}}, \bibinfo{pages}{4} (\bibinfo{year}{2007}).



\bibitem[{\citenamefont{Franceschetti and Zunger}(1999)}]{Franceschetti}
\bibinfo{author}{\bibfnamefont{A.}~\bibnamefont{Franceschetti}}
  \bibnamefont{and} \bibinfo{author}{\bibfnamefont{A.}~\bibnamefont{Zunger}},
  \bibinfo{journal}{Nature} \textbf{\bibinfo{volume}{402}}, \bibinfo{pages}{60}
  (\bibinfo{year}{1999}).

\bibitem[{\citenamefont{Car and Parrinello}(1985)}]{CarParrinello}
\bibinfo{author}{\bibfnamefont{R.}~\bibnamefont{Car}} \bibnamefont{and}
  \bibinfo{author}{\bibfnamefont{M.}~\bibnamefont{Parrinello}},
  \bibinfo{journal}{Phys. Rev. Lett.} \textbf{\bibinfo{volume}{55}},
  \bibinfo{pages}{2471} (\bibinfo{year}{1985}).

\bibitem[{\citenamefont{Bruno et~al.}(2008)\citenamefont{Bruno, Mammano,
  Fiorino, and Morabito}}]{BMFM}
\bibinfo{author}{\bibfnamefont{E.}~\bibnamefont{Bruno}},
  \bibinfo{author}{\bibfnamefont{F.}~\bibnamefont{Mammano}},
  \bibinfo{author}{\bibfnamefont{A.}~\bibnamefont{Fiorino}}, \bibnamefont{and}
  \bibinfo{author}{\bibfnamefont{E.~V.} \bibnamefont{Morabito}},
  \bibinfo{journal}{Phys. Rev. B} \textbf{\bibinfo{volume}{77}},
  \bibinfo{eid}{155108} (\bibinfo{year}{2008}).



\bibitem[{not({\natexlab{a}})}]{nota1}
\bibinfo{note}{In Eq.~(\ref{eq1}) and throughout this Letter, the summations
  over the angular momentum components are truncated at $\ell=0$ for sake of
  simplicity.}

\bibitem[{\citenamefont{Dreizler and Gross}(1990)}]{Dreizler&Gross}
\bibinfo{author}{\bibfnamefont{R.~M.} \bibnamefont{Dreizler}} \bibnamefont{and}
  \bibinfo{author}{\bibfnamefont{E.~K.~U.} \bibnamefont{Gross}},
  \emph{\bibinfo{title}{Density Functional Theory}}
  (\bibinfo{publisher}{Springer-Verlag, Berlin}, \bibinfo{year}{1990}).

\bibitem[{\citenamefont{Gonis}(1992)}]{Gonis}
\bibinfo{author}{\bibfnamefont{A.}~\bibnamefont{Gonis}},
  \emph{\bibinfo{title}{Green Functions for Ordered and Disordered Systems}}
  (\bibinfo{publisher}{North-Holland, Amsterdam}, \bibinfo{year}{1992}).

\bibitem[{\citenamefont{Harris}(1985)}]{Harris}
\bibinfo{author}{\bibfnamefont{J.}~\bibnamefont{Harris}},
  \bibinfo{journal}{Phys. Rev. B} \textbf{\bibinfo{volume}{31}},
  \bibinfo{pages}{1770} (\bibinfo{year}{1985}).

\bibitem[{\citenamefont{Krajewski and Parrinello}(2005)}]{Krajewski&Parrinello}
\bibinfo{author}{\bibfnamefont{F.~R.} \bibnamefont{Krajewski}}
  \bibnamefont{and}
  \bibinfo{author}{\bibfnamefont{M.}~\bibnamefont{Parrinello}},
  \bibinfo{journal}{Phys. Rev. B} \textbf{\bibinfo{volume}{71}},
  \bibinfo{pages}{233105} (\bibinfo{year}{2005}).

\bibitem[{\citenamefont{Soven}(1967)}]{Soven}
\bibinfo{author}{\bibfnamefont{P.}~\bibnamefont{Soven}},
  \bibinfo{journal}{Phys. Rev.} \textbf{\bibinfo{volume}{156}},
  \bibinfo{pages}{809} (\bibinfo{year}{1967}).

\bibitem[{\citenamefont{Abrikosov and Johansson}(1998)}]{Abrikosov_cpa}
\bibinfo{author}{\bibfnamefont{I.~A.} \bibnamefont{Abrikosov}}
  \bibnamefont{and}
  \bibinfo{author}{\bibfnamefont{B.}~\bibnamefont{Johansson}},
  \bibinfo{journal}{Phys. Rev. B} \textbf{\bibinfo{volume}{57}},
  \bibinfo{pages}{14164} (\bibinfo{year}{1998}).

\bibitem[{\citenamefont{Magri et~al.}(1990)\citenamefont{Magri, Wei, and
  Zunger}}]{Magri}
\bibinfo{author}{\bibfnamefont{R.}~\bibnamefont{Magri}},
  \bibinfo{author}{\bibfnamefont{S.~H.} \bibnamefont{Wei}}, \bibnamefont{and}
  \bibinfo{author}{\bibfnamefont{A.}~\bibnamefont{Zunger}},
  \bibinfo{journal}{Phys. Rev. B} \textbf{\bibinfo{volume}{42}},
  \bibinfo{pages}{11388} (\bibinfo{year}{1990}).

\bibitem[{\citenamefont{Ujfalussy et~al.}(2000)\citenamefont{Ujfalussy,
  Faulkner, Moghadam, Stocks, and Wang}}]{Ujfalussy}
\bibinfo{author}{\bibfnamefont{B.}~\bibnamefont{Ujfalussy}}~et~al.,
  \bibinfo{journal}{Phys. Rev. B} \textbf{\bibinfo{volume}{61}},
  \bibinfo{pages}{12005} (\bibinfo{year}{2000}).



\bibitem[{\citenamefont{Johnson et~al.}(1986)\citenamefont{Johnson, Nicholson,
  Pinsky, Gyorffy, and Stocks}}]{DFTKKRCPA1}
\bibinfo{author}{\bibfnamefont{D.~D.} \bibnamefont{Johnson}}~et~al.,
\bibinfo{journal}{Phys. Rev. Lett.}
  \textbf{\bibinfo{volume}{56}}, \bibinfo{pages}{2088} (\bibinfo{year}{1986}).



\bibitem[{\citenamefont{Johnson et~al.}(1990)\citenamefont{Johnson, Nicholson,
  Pinsky, Gyorffy, and Stocks}}]{DFTKKRCPA2}
\bibinfo{author}{\bibfnamefont{D.~D.} \bibnamefont{Johnson}}~et~al.,
  \bibinfo{journal}{Phys. Rev. B}
  \textbf{\bibinfo{volume}{41}}, \bibinfo{pages}{9701} (\bibinfo{year}{1990}).

\bibitem[{\citenamefont{Faulkner et~al.}(1995)\citenamefont{Faulkner, Wang, and
  Stocks}}]{FWS1}
\bibinfo{author}{\bibfnamefont{J.~S.} \bibnamefont{Faulkner}},
  \bibinfo{author}{\bibfnamefont{Y.}~\bibnamefont{Wang}}, \bibnamefont{and}
  \bibinfo{author}{\bibfnamefont{G.~M.} \bibnamefont{Stocks}},
  \bibinfo{journal}{Phys. Rev. B} \textbf{\bibinfo{volume}{52}},
  \bibinfo{pages}{17106} (\bibinfo{year}{1995}).

\bibitem[{\citenamefont{Bruno et~al.}(2003)\citenamefont{Bruno, Zingales, and
  Wang}}]{CEF}
\bibinfo{author}{\bibfnamefont{E.}~\bibnamefont{Bruno}},
  \bibinfo{author}{\bibfnamefont{L.}~\bibnamefont{Zingales}}, \bibnamefont{and}
  \bibinfo{author}{\bibfnamefont{Y.}~\bibnamefont{Wang}},
  \bibinfo{journal}{Phys. Rev. Lett.} \textbf{\bibinfo{volume}{91}},
  \bibinfo{pages}{166401} (\bibinfo{year}{2003}).

\bibitem[{\citenamefont{Bruno}(2007)}]{brunomatsci}
\bibinfo{author}{\bibfnamefont{E.}~\bibnamefont{Bruno}}, \bibinfo{journal}{Mat.
  Sci. and Engin. A} \textbf{\bibinfo{volume}{462}}, \bibinfo{pages}{456}
  (\bibinfo{year}{2007}).

\bibitem[{\citenamefont{Metropolis et~al.}(1953)\citenamefont{Metropolis,
  Rosenbluth, Rosenbluth, Teller, and Teller}}]{Metropolis}
\bibinfo{author}{\bibfnamefont{M.}~\bibnamefont{Metropolis}}~et~al.,
  \bibinfo{journal}{J. Chem. Phys.} \textbf{\bibinfo{volume}{21}},
  \bibinfo{pages}{1087} (\bibinfo{year}{1953}).


\bibitem[{\citenamefont{Landau and Binder}(2000)}]{Landau&Binder}
\bibinfo{author}{\bibfnamefont{D.~P.} \bibnamefont{Landau}} \bibnamefont{and}
  \bibinfo{author}{\bibfnamefont{K.}~\bibnamefont{Binder}},
  \emph{\bibinfo{title}{A Guide to Monte Carlo Simulations in Statistical
  Physics}} (\bibinfo{publisher}{Cambridge University Press},
  \bibinfo{year}{2000}).

\bibitem[{not({\natexlab{b}})}]{nota2}
\bibinfo{note}{See also the EPAPS accompanying the present Letter.}

\bibitem[{not({\natexlab{c}})}]{nota3}
\bibinfo{note}{We have performed PCPA calculations for unit cells containing a
  few hundreds atoms, the atomic sphere approximation for the site potentials
  and the LDA treatment for the exchange-correlation functional.}

\bibitem[{\citenamefont{Muller and Zunger}(2001)}]{alphabrass}
\bibinfo{author}{\bibfnamefont{S.}~\bibnamefont{Muller}} \bibnamefont{and}
  \bibinfo{author}{\bibfnamefont{A.}~\bibnamefont{Zunger}},
  \bibinfo{journal}{Phys. Rev. B} \textbf{\bibinfo{volume}{63}},
  \bibinfo{pages}{094204} (\bibinfo{year}{2001}).

\bibitem[{\citenamefont{Tarafder et~al.}(2006)\citenamefont{Tarafder,
  Chakrabarti, Saha, and Mookerjee}}]{optcuzn}
\bibinfo{author}{\bibfnamefont{K.}~\bibnamefont{Tarafder}},
  \bibinfo{author}{\bibfnamefont{A.}~\bibnamefont{Chakrabarti}},
  \bibinfo{author}{\bibfnamefont{K.~K.} \bibnamefont{Saha}}, \bibnamefont{and}
  \bibinfo{author}{\bibfnamefont{A.}~\bibnamefont{Mookerjee}},
  \bibinfo{journal}{Phys. Rev. B} \textbf{\bibinfo{volume}{74}},
  \bibinfo{pages}{144204} (\bibinfo{year}{2006}).

\bibitem[{\citenamefont{Tulip et~al.}(2006)\citenamefont{Tulip, Staunton,
  Rowlands, Gyorffy, Bruno, and Ginatempo}}]{Tulip}
\bibinfo{author}{\bibfnamefont{P.~R.} \bibnamefont{Tulip}}~et~al.,
  \bibinfo{journal}{Phys. Rev. B} \textbf{\bibinfo{volume}{73}},
  \bibinfo{pages}{205109} (\bibinfo{year}{2006}).

\bibitem[{\citenamefont{Landau and Lifshitz}(1980)}]{LLstphysI}
\bibinfo{author}{\bibfnamefont{L.~D.} \bibnamefont{Landau}} \bibnamefont{and}
  \bibinfo{author}{\bibfnamefont{E.~M.} \bibnamefont{Lifshitz}},
  \emph{\bibinfo{title}{Statistical Physics, Part I}}
  (\bibinfo{publisher}{Pergamon Press}, \bibinfo{address}{Oxford},
  \bibinfo{year}{1980}).

\bibitem[{\citenamefont{Johnson et~al.}(2000)\citenamefont{Johnson, Smirnov,
  Staunton, Pinski, and Shelton}}]{Johnson2000}
\bibinfo{author}{\bibfnamefont{D.~D.} \bibnamefont{Johnson}}~et~al.,
  \bibinfo{journal}{Phys. Rev. B} \textbf{\bibinfo{volume}{62}},
  \bibinfo{pages}{R11917} (\bibinfo{year}{2000}).

\bibitem[{\citenamefont{Zarkevich and Johnson}(2004)}]{Zarkevich2004}
\bibinfo{author}{\bibfnamefont{N.~A.} \bibnamefont{Zarkevich}}
  \bibnamefont{and} \bibinfo{author}{\bibfnamefont{D.~D.}
  \bibnamefont{Johnson}}, \bibinfo{journal}{Phys. Rev. Lett.}
  \textbf{\bibinfo{volume}{92}}, \bibinfo{pages}{255702}
  (\bibinfo{year}{2004}).

\bibitem[{\citenamefont{Moruzzi et~al.}(1988)\citenamefont{Moruzzi, Janak, and
  Schwartz}}]{DebyeGruneisen}
\bibinfo{author}{\bibfnamefont{V.~L.} \bibnamefont{Moruzzi}},
  \bibinfo{author}{\bibfnamefont{J.}~\bibnamefont{Janak}}, \bibnamefont{and}
  \bibinfo{author}{\bibfnamefont{K.}~\bibnamefont{Schwartz}},
  \bibinfo{journal}{Phys. Rev. B} \textbf{\bibinfo{volume}{37}},
  \bibinfo{pages}{790} (\bibinfo{year}{1988}).

\bibitem[{ASM(1992)}]{ASM}
\emph{\bibinfo{title}{ASM Handbook, vol. 3, Alloy Phase Diagrams}}
  (\bibinfo{year}{1992}).

\end{thebibliography}

\end{document}